\documentclass[aps,prd,preprint,superscriptaddress,nofootinbib,amsmath,amssymb]{revtex4}
\usepackage{graphicx}
\usepackage{dcolumn}
\usepackage{bm}
\usepackage{amsmath}
\usepackage{amsfonts}
\usepackage{amssymb}
\usepackage{appendix}
\usepackage{pdfpages}
\usepackage{graphicx}
\usepackage{wrapfig}
\usepackage{multirow}
\usepackage{mathrsfs}
\usepackage{epstopdf}
\usepackage{slashed}
\usepackage{soul}
\usepackage{mathtools}
\usepackage{floatrow}
\usepackage{float}
\usepackage{csquotes}
\usepackage{wrapfig}
\usepackage{epstopdf}
\usepackage{graphicx}
\usepackage{caption}
\usepackage{epsfig}
\usepackage{hyperref}
\usepackage[utf8]{inputenc}
\usepackage{epsfig}
\usepackage{dcolumn}
\usepackage{morefloats}

\def\be{\begin{equation}}
\def\ee{\end{equation}}
\def\bea{\begin{eqnarray}}
\def\eea{\end{eqnarray}}
\def\gsim{\ \rlap{\raise 2pt\hbox{$>$}}{\lower 2pt \hbox{$\sim$}}\ }
\def\lsim{\ \rlap{\raise 2pt\hbox{$<$}}{\lower 2pt \hbox{$\sim$}}\ }
\def\dslash{\kern-4pt \not{\hbox{\kern-2pt $\partial$}}}
\def\pslash{\not{\hbox{\kern-2pt p}}}

\newcommand{\dcp}{\delta_{CP}}

\begin{document}
\DeclareGraphicsExtensions{.eps,.ps}
\title{\boldmath Challenges posed by non-standard neutrino interactions in the determination of $\delta_{CP}$ at DUNE}
\author{K. N. Deepthi}
\email[Email Address: ]{deepthi@prl.res.in}
\affiliation{
Physical Research Laboratory, Navrangpura,
Ahmedabad 380 009, India}

\author{Srubabati Goswami}
\email[Email Address: ]{sruba@prl.res.in}
\affiliation{
Physical Research Laboratory, Navrangpura,
Ahmedabad 380 009, India}

\author{Newton Nath}  
\email[Email Address: ]{newton@ihep.ac.cn}
\affiliation{Institute of High Energy Physics, Chinese Academy of Sciences, Beijing-100049, China.}

\begin{abstract} 
One of the primary objectives of Deep Underground Neutrino Experiment (DUNE) is to discover the leptonic CP violation and to identify it's source. 
In this context, we study the impact of non-standard neutrino interactions (NSIs) on observing the CP violation signal at DUNE. 
We explore the impact of various parameter degeneracies introduced by non-zero NSI and identify which of these can influence the CP violation sensitivity
and CP precision of DUNE, by considering NSI both in data and in theory. 
In particular, we study how the CP sensitivity of DUNE  is affected because
of the intrinsic hierarchy degeneracy which occurs when the diagonal 
NSI parameter $\epsilon_{ee}=-1$ and $\delta_{CP}= \pm 90^{\circ}$.
\end{abstract}
\maketitle
\textbf{Introduction :} Neutrino oscillation physics has entered an era filled with series of 
exceptional ongoing and forthcoming experiments using diverse sources and various detection techniques. 
These experiments are aimed at addressing some pressing questions like 
the determination of CP violation in the leptonic sector, 
ordering of neutrino masses i.e whether they obey 
normal hierarchy (NH, $m_3 > m_2 > m_1$) or 
inverted hierarchy (IH, $m_3 < m_1 \approx m_2$), 
octant of $\theta_{23}$ i.e  whether it falls in the 
lower octant (LO, $\theta_{23} < \pi/4$) or in the higher octant (HO, $\theta_{23} > \pi/4$).

Numerous neutrino experiments have left a window open for several new physics scenarios among which non-standard neutrino interactions (NSIs) have received a lot of attention lately. A detailed review on NSIs can be found in Refs.\cite{Ohlsson:2012kf, Miranda:2015dra, Farzan:2017xzy}. NSIs were first proposed by Wolfenstein \cite{Wolfenstein:1977ue} as an alternative phenomenon to explain neutrino flavour transitions. 
 However, experiments over past decades have established neutrino oscillations as the leading mechanism to flavor transitions leaving non-standard interactions of neutrinos with matter particles as potential next to leading order effects. Precision measurements from ongoing and future neutrino experiments could provide some insight into the existence of these new type of interactions. Among these, DUNE (Deep Underground Neutrino Experiment) is considered as one of the powerful experiments \cite{Acciarri:2016crz}. Therefore, it is important to phenomenologically understand the effect of these beyond SM interactions on the experimental sensitivities of DUNE. 

DUNE is a long baseline accelerator neutrino experiment which will use very high intensity neutrinos produced by proposed Long Baseline Neutrino Facility's (LBNF) beamline at Fermilab, USA \cite{Acciarri:2015uup}. Fermilab's main injector accelerator provides a proton beam corresponding to an energy of 120 GeV which in turn generates a broad band neutrino beam of energy ranging from 100 MeV to 8 GeV with a peak at 2.5 GeV. This neutrino beam is then made to travel 1300 km towards Sanford Underground Research Facility (SURF) where it encounters a Liquid Argon Time-Projection Chamber (LArTPC). We consider a detector of 40 kt fiducial volume in our simulations. 

CP violation in the leptonic sector holds a key to leptogenesis which in turn can shed some light on baryogenesis \cite{Fukugita:1986hr,Davidson:2008bu}. Several phases could contribute to leptonic CP violation. Among these - Dirac CP violation phase $\delta_{CP}$ is one of the physical phases (apart from two Majorana phases) occurring in neutrino mixing matrices which can be determined via long baseline (LBL) neutrino oscillation experiments. Recent global analysis of neutrino oscillation measurements \cite{Capozzi:2016rtj,Esteban:2016qun} have suggested $\delta_{CP}$ to be $-90^\circ$ and have shown a weak trend towards NH. However, these experiments suffer from various parameter degeneracies because of their shorter baselines as compared to DUNE \cite{Barger:2001yr,Ghosh:2015ena, Goswami:2017hcw}. Furthermore,
given the potency of upcoming LBL neutrino oscillation experiments, like DUNE \cite{Acciarri:2015uup} and Hyper-Kamiokande \cite{hkloi}, it is a feasible task to determine $\delta_{CP}$ with high sensitivity. Additionally, there is a possibility that these experiments will be able to probe NSI and further identify any new sources of CP-violation. Recent studies considering the impact of NSI in the context of DUNE can be found in \cite{Masud:2015xva,deGouvea:2015ndi,Coloma:2015kiu,Liao:2016hsa,Soumya:2016enw,Blennow:2016etl,Forero:2016cmb,Huitu:2016bmb,
 Masud:2016bvp,Coloma:2016gei,Masud:2016gcl,Agarwalla:2016fkh,
Liao:2016bgf,Fukasawa:2016gvm,Blennow:2016jkn,Liao:2016orc,
Deepthi:2016erc,Ghosh:2017ged,Ghosh:2017lim,Masud:2017bcf} and the references there in. Below is the brief account of 
the papers which dealt with the effect of NSI on the CP violation sensitivity of DUNE.

In Ref. \cite{Masud:2015xva} the authors have studied the effect of non-zero complex NSI parameters on the CP violation sensitivity of DUNE and showed that these complex phases could mimic CP violation even for the CP conserving values. The authors of Ref. \cite{deGouvea:2015ndi} have investigated how well DUNE can measure the new sources of CP violation assuming the presence of NSI. 
Ref. \cite{Liao:2016hsa} discusses several parameter degeneracies between standard and non-standard interactions and the effect on the determination of mass hierarchy, octant of $\theta_{23}$ and the CPV sensitivity of DUNE. Authors of Ref. \cite{Forero:2016cmb} have tested the robustness of the recent result of Dirac CP phase $\dcp=-90^\circ$ by assuming NSI in the simulated data (generated for many iterations) and fitting it with SI. While Ref. \cite{Masud:2016bvp} deals with precise study of the impact of NSI on the CP violation sensitivity of the long-baseline experiments like T2K, NO$\nu$A, DUNE and T2HK. In Ref. \cite{Huitu:2016bmb} the authors have constrained the NSI parameters while considering standard interactions in the data and non-standard interactions in theory. In Ref. \cite{deGouvea:2016pom}, authors have investigated false CPV 
signals from new physics scenarios like NSI and a four-neutrino scenario at DUNE. Ref. \cite{Liao:2016orc} investigates the effect of matter NSI on the determination of Dirac CP phase at DUNE, T2HK and T2HKK while studying the role played by the generalized mass hierarchy degeneracy introduced by $\epsilon_{ee} \rightarrow -\epsilon_{ee}-2$ \cite{Bakhti:2014pva,Coloma:2016gei}. 


In this work we investigate how matter induced non-standard neutrino interactions would affect the determination of CP violation by DUNE, during its proposed run time of (5 $\nu$ + 5 $\bar{\nu}$) years. 
We focus on the impact of various parameter degeneracies, introduced by the non-standard neutrino interactions. We consider the most general degeneracy of the form $(\epsilon_{ee}, \dcp) \rightarrow (\epsilon_{ee}^\prime, \dcp^\prime)$ occurring for same as well as opposite hierarchies  
and study how this can affect the CP sensitivity of DUNE. 
In particular, 
we study the effect of the generalized mass ordering degeneracy $\epsilon_{ee}\rightarrow -\epsilon_{ee}-2$, $\dcp \rightarrow \pi - \dcp$ stressing on the special case of $\epsilon_{ee}= -1$ and $\delta_{CP}= \pm 90^{\circ}$ assuming NSI both in data and in theory. This particular case is interesting as $\epsilon_{ee} = -1$ nullifies the standard matter effect \footnote{Note that though $\epsilon_{ee} = -1$ cancels the matter effect in the Earth it doesn't cancel the matter effect in the Sun \cite{Deepthi:2016erc}. For the impact of large $\epsilon_{ee}$ in the solar matter effect see \cite{Miranda:2004nb}}. Additionally if there is maximal CP violation $\delta_{CP}= \pm 90^{\circ}$ there exists an intrinsic hierarchy degeneracy that is independent of baseline and neutrino beam energy \cite{Deepthi:2016erc}. We study whether the intrinsic degeneracy can 
alter the CP discovery potential and CP precision of DUNE. 
In addition, we consider other true values of $\epsilon_{ee} \in [-4,4]$ and obtain the fraction of $\delta_{CP}$ values for which CP violation (CPV) can be discovered at $3\sigma$. We also 
investigate the impact of introducing non-zero value of the 
off-diagonal NSI parameter $\epsilon_{e \tau}$ along with $\epsilon_{ee} = -1$.

We organize the paper as follows: We present a brief account of NSI formalism and their current bounds in section \ref{sec:NSI form}. In section \ref{sec:prob}, we present a discussion on the probability and CP asymmetry of DUNE. In section \ref{sec:CPV sens} we discuss the effect of diagonal and off-diagonal NSI on the CPV discovery capability of DUNE assuming the presence of NSI in both data and theory. Section \ref{sec:cp_precision}, discusses the impact of NSI on the CP precision of DUNE. Concluding remarks are made in section \ref{sec:conclusions}.
\section{NSI formalism and current bounds} \label{sec:NSI form}
 Neutral current NSIs that affect the propagation of neutrinos through earth matter are described using an effective four-fermion Lagrangian with dimension $d = 6$ operators of the form
\be
  \label{eq:NSI}
  \mathcal{L}^{NC}_\text{NSI} = -2\sqrt{2}G_F
   \sum_{f,\alpha,\beta,C} \epsilon^{fC}_{\alpha\beta}(\overline{\nu}_\alpha \gamma^{\mu} P_{L} \nu_\beta )
        ( \bar{f} \gamma_{\mu} P_C f ) + \text{h.c.}
\ee
where $\epsilon^{fC}_{\alpha\beta}$ are NSI parameters, $\alpha, \beta = e, \mu, \tau$, $C = L,R$, $f = u,d,e$, and $ G_{F} $ is the Fermi constant. 
The Schrödinger-like evolution equation of a neutrino $\nu = (\nu_{e}~\nu_{\mu}~\nu_{\tau})^{T}$, travelling a distance x, can be written as  
\begin{widetext}
\begin{eqnarray} 
\hspace{-2.7cm}
i {\frac{d\nu}{dx}}
=  H^{\prime} \nu
\label{eq:general-evolution}
\end{eqnarray}
where the effective Hamiltonian $H^{\prime}$ characterizes both standard and non-standard interactions of neutrinos with matter fermions and can be  expressed
in the flavour basis as,  
\begin{eqnarray}
H^{\prime} = \frac{1}{2E} \left\{
U \left[ \begin{array}{ccc}
       0   & 0          & 0   \\
       0   &\Delta_{21} & 0  \\
       0   & 0          &\Delta_{31}  
\end{array} \right]U^{\dagger} 
 +  A \left(\begin{array}{ccc}
1 + \epsilon_{ee} & \epsilon_{e\mu}e^{i\phi_{e\mu}} & \epsilon_{e\tau}e^{i\phi_{e\tau}}
\\
\epsilon_{e\mu}e^{-i\phi_{e\mu}} & \epsilon_{\mu\mu} & \epsilon_{\mu\tau}e^{i\phi_{\mu\tau}}
\\
\epsilon_{e\tau}e^{-i\phi_{e\tau}}& \epsilon_{\mu\tau}e^{-i\phi_{\mu\tau}} & \epsilon_{\tau\tau}
\end{array}\right)\,
\right\}
\label{eq:Hamiltonian}
\end{eqnarray}
\end{widetext}
The `1' in the effective matter potential of eq.~(\ref{eq:Hamiltonian}) corresponds to the standard charged-current matter interactions whereas, the NSI parameters $\epsilon_{\alpha\beta}$ describe the non-standard interactions of neutrinos with earth matter.
Here, $U$ is the Pontecorvo-Maki-Nakagawa-Sakata (PMNS) mixing matrix~\cite{Patrignani:2016xqp}, $\Delta_{ij}=m^2_i-m^2_j$, $A \equiv 2\sqrt2 G_F N_e E$ and $\epsilon_{\alpha\beta}e^{i\phi_{\alpha\beta}}\equiv\sum\limits_{f,C}\epsilon^{f C}_{\alpha\beta}\frac{N_f}{N_e}$, with $N_f$ being the number density of fermion $f$. 

Currently available conservative model independent bounds \cite{Blennow:2014sja,Ohlsson:2012kf} on NSI parameters are given as, 
\begin{eqnarray}
 \left(\begin{array}{ccc}
 |\epsilon_{ee}|<4.2 & |\epsilon_{e\mu}|<0.33 & |\epsilon_{e\tau}|<3.0
\\
 & |\epsilon_{\mu\mu}|<0.07 & |\epsilon_{\mu\tau}|<0.33
\\
&  & |\epsilon_{\tau\tau}|<21
\end{array}\right)\,
\end{eqnarray}
%
\section{Probability and CP asymmetry for DUNE :}\label{sec:prob}
The relevant oscillation probability for the super beam experiment DUNE is the appearance channel ($P_{\mu e}$). This can be expressed in terms of small parameters, $s_{13}$, $r = \Delta m^2_{21}/\Delta m^2_{31}$ and $\epsilon_{\alpha \beta}$ except $ \alpha, \beta = e$ for normal hierarchy (NH) as \cite{Liao:2016hsa}
\begin{widetext}
\bea
P_{\mu e} &=& x^2 f^2 + 2xyfg \cos(\Delta + \delta_{CP}) + y^2 g^2
\nonumber\\
&+& 4\hat A \epsilon_{e\mu}
\left\{ xf [s_{23}^2 f \cos(\phi_{e\mu}+\delta)  
+ c_{23}^2 g \cos(\Delta+\delta+\phi_{e\mu})]\right.
 \left. +yg [c_{23}^2 g \cos\phi_{e\mu} + s_{23}^2 f \cos(\Delta-\phi_{e\mu})]\right\}
\nonumber\\
&+& 4\hat A \epsilon_{e\tau} s_{23} c_{23}
\left\{ xf [f \cos(\phi_{e\tau}+\delta)  
- g \cos(\Delta+\delta+\phi_{e\tau})] \right.
 \left. -yg [g \cos\phi_{e\tau} - f \cos(\Delta-\phi_{e\tau})]\right\}
\nonumber\\
&+& 4 \hat A^2 g^2 c_{23}^2 |c_{23} \epsilon_{e\mu} - s_{23}\epsilon_{e\tau}|^2
 +  4 \hat A^2 f^2 s_{23}^2 |s_{23} \epsilon_{e\mu} + c_{23}\epsilon_{e\tau}|^2
\nonumber\\
&+& 8 \hat A^2 fg s_{23} c_{23}
\left\{ c_{23}\cos\Delta
\left[ s_{23}(\epsilon_{e\mu}^2 - \epsilon_{e\tau}^2)\right.\right.
 \left.\left.+2 c_{23} \epsilon_{e\mu}\epsilon_{e\tau}
\cos(\phi_{e\mu}-\phi_{e\tau})\right]\right.
  \left.-\epsilon_{e\mu}\epsilon_{e\tau}
\cos(\Delta-\phi_{e\mu}+\phi_{e\tau})\right\}
\nonumber\\
&+& {\cal O}(s_{13}^2 \epsilon, s_{13}\epsilon^2, \epsilon^3)\,,
\label{eq:prob}
\eea
where
\bea
&& x = 2 s_{13} s_{23},~
y = 2r s_{12} c_{12} c_{23},~
 (s_{ij} = \sin\theta_{ij},c_{ij}=\cos\theta_{ij}, i<j,i,j=1,2,3 ) 
\nonumber\\
\Delta &=& \frac{\Delta m^2_{31} L}{4E},\ 
\hat A = \frac{A}{\Delta m^2_{31}},\
f,\, \bar{f} = \frac{\sin[\Delta(1\mp\hat A(1+\epsilon_{ee}))]}{(1\mp\hat A(1+\epsilon_{ee}))}\,,\ 
g = \frac{\sin[\hat A(1+\epsilon_{ee}) \Delta]}{\hat A(1+\epsilon_{ee})}
\label{eq:define}
\eea
\end{widetext}
The expressions for the inverted hierarchy (IH) can be obtained by replacing 
$\Delta m^2_{31} \to - \Delta m^2_{31}$ (i.e.  $ \Rightarrow \Delta \to - \Delta$,  $\hat A \to - \hat A$ 
(i.e. $f \to - \bar{f}$ and $g \to -g$), $ y \to -y $ ). 
Similar expressions for  antineutrino probability 
($ P_{\overline{\mu} \overline{e}} $) can be obtained by replacing 
 $\hat A \to - \hat A$ ( $ \Rightarrow f \to \bar{f}$),
$\delta_{CP} \to - \delta_{CP}$. 
We note from the above that the 
only diagonal parameter to which appearance channel is sensitive to is
$ \epsilon_{ee} $. In the absence of off-diagonal NSI parameters, the probability expression is 
just the first line of eq.(\ref{eq:prob}). The  NSI contribution appears only in $f,~ \bar{f} $ and  $ g $ terms via $ \epsilon_{ee} $ which is not treated as a small parameter in this formulation.

\begin{figure}[h!]
        \begin{tabular}{lr}
                \hspace*{-0.35in} 
                \includegraphics[height=6cm,width=7cm]{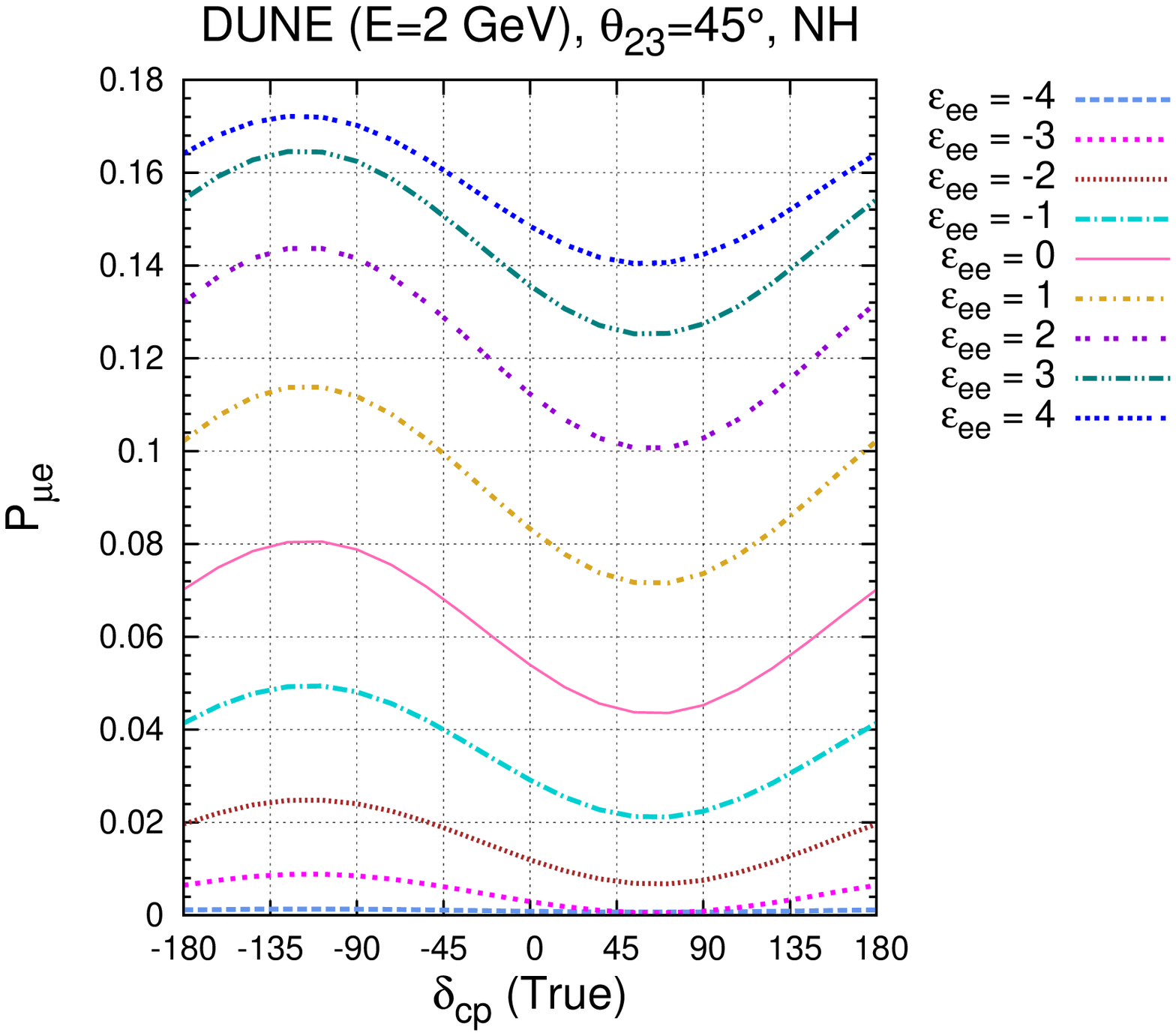} 
                \includegraphics[height=6cm,width=6cm]{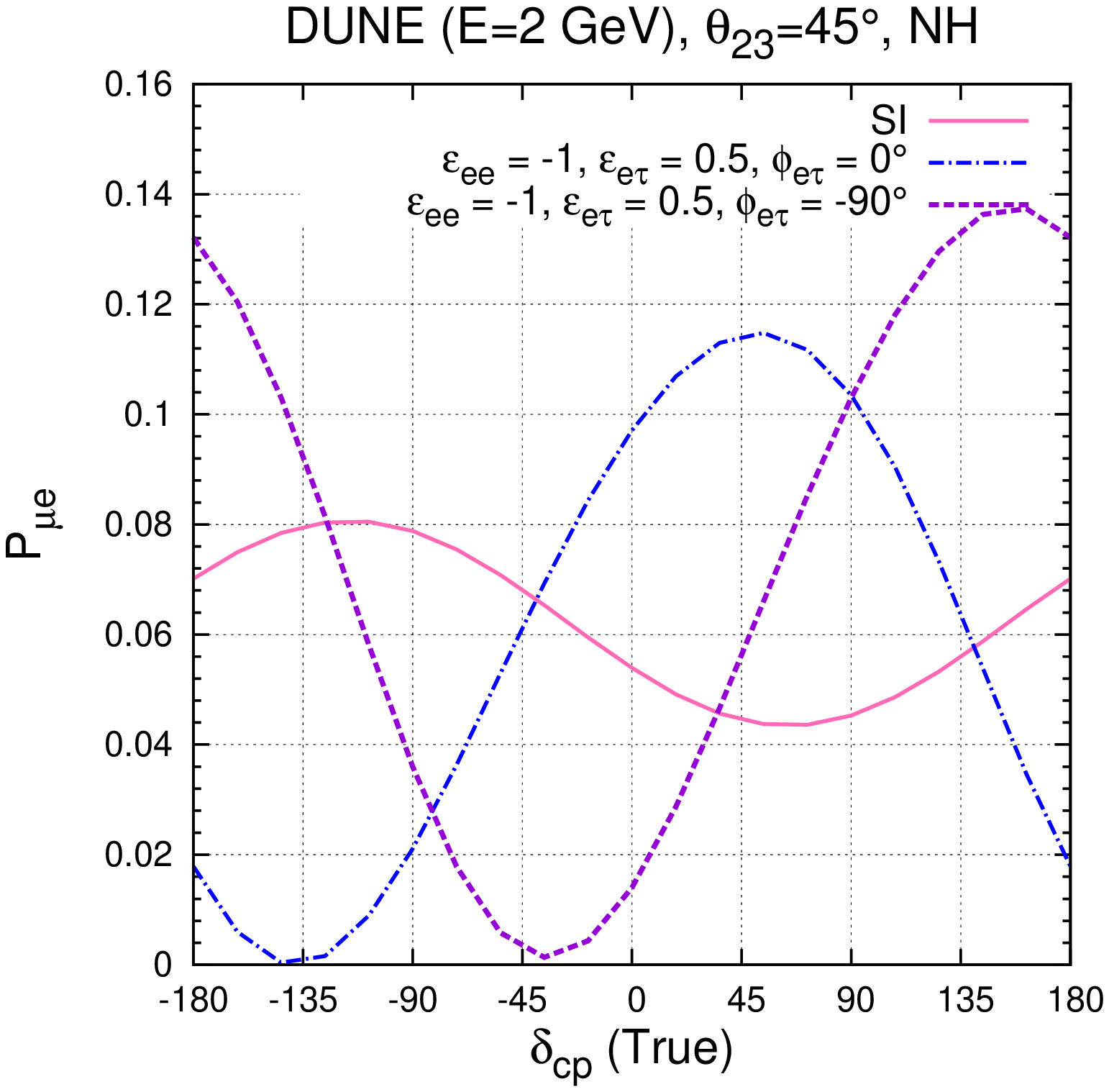}
        \end{tabular}
\vspace{-0.7cm}        
\caption{\footnotesize Left Panel: $ P_{\mu e}  $  vs $ \delta_{CP} $ for DUNE for energy (E = 2 GeV) assuming the presence of non-zero diagonal NSI parameter $\epsilon_{ee}$. Right Panel: $ P_{\mu e}  $  vs $ \delta_{CP} $ assuming the $\epsilon_{ee} = -1$ and non-zero complex off-diagonal NSI parameter $\epsilon_{e \tau}$.}
\label{fig:WH-WCP}
\end{figure}

 Fig. (\ref{fig:WH-WCP}) shows $ P_{\mu e}  $  vs $ \delta_{CP} $ of DUNE for E = 2 GeV with $\theta_{23} = 45^\circ$ and assuming normal hierarchy. The plot in the left panel shows $ P_{\mu e}  $ as a function of $ \delta_{CP} $ for different values of $\epsilon_{ee} \in [-4,4]$. 
The standard case without NSI corresponds to the pink solid curve with $\epsilon_{ee} = 0$. 
It can be seen that there exists degeneracies of the form ($\epsilon \rightarrow \epsilon^\prime$, $\dcp \rightarrow \dcp^\prime$) even for the same hierarchy. This can be understood by drawing a horizontal line intersecting any two curves. Moreover, if one plots $ P_{\mu e}$ vs $ \delta_{CP} $ for all non-integral values of $\epsilon_{ee} \in [-4,4]$ there will be a continous degeneracy of this form which effects the determination of CP phase at DUNE. Note that this kind of degeneracies between opposite hierarchies have already been elaborated in \cite{Deepthi:2016erc}.
 
 In the right panel we show the effect of non-zero $\epsilon_{e \tau}$ on $ P_{\mu e}  $ for $\epsilon_{ee} = -1$. We consider a representative value of $\epsilon_{e \tau} = 0.5$ and $\phi_{e \tau} = 0^\circ, -90^\circ$ to see the effect of complex diagonal NSI\footnote{In this paper we study the effect of only non-zero $\epsilon_{e \tau}$ and not $\epsilon_{e \mu}$ as the latter has tighter bounds.}.
 It can be seen that the standard case (no NSI) shown by pink (solid) curve is degenerate with the non-zero NSI case represented by blue (dot-dashed)
and the violet (dotted) curves for four different values of $\delta_{CP}$. 
Apart from that, one can also find many degeneracies when a horizontal line is drawn intersecting the lines. These degeneracies correspond to wrong $\dcp$ solutions which play a role in CPV sensitivity of DUNE if NSI exists in nature (as shown in the next section) even when the hierarchy is known.
%
%

In neutrino oscillation experiments the CP-violating effects can be characterized by a quantity known as 
CP asymmetry which is defined as $A_{CP}= P_{\alpha \beta} - \bar{P}_{\alpha \beta}$ ($\alpha \neq \beta$). In the case of vacuum oscillations $A_{CP} \propto \sin \delta$ and clearly the Dirac CP phase $\dcp$
can be easily determined. However for long baseline experiments like DUNE large matter effects induce fake CP phase ($\dcp^{'}$) which cannot be distinguished from the intrinsic CP phase ($\dcp$).

\begin{figure}[h!]
        \begin{tabular}{lr}
                \hspace*{-0.35in} 
                \includegraphics[height=6cm,width=7cm]{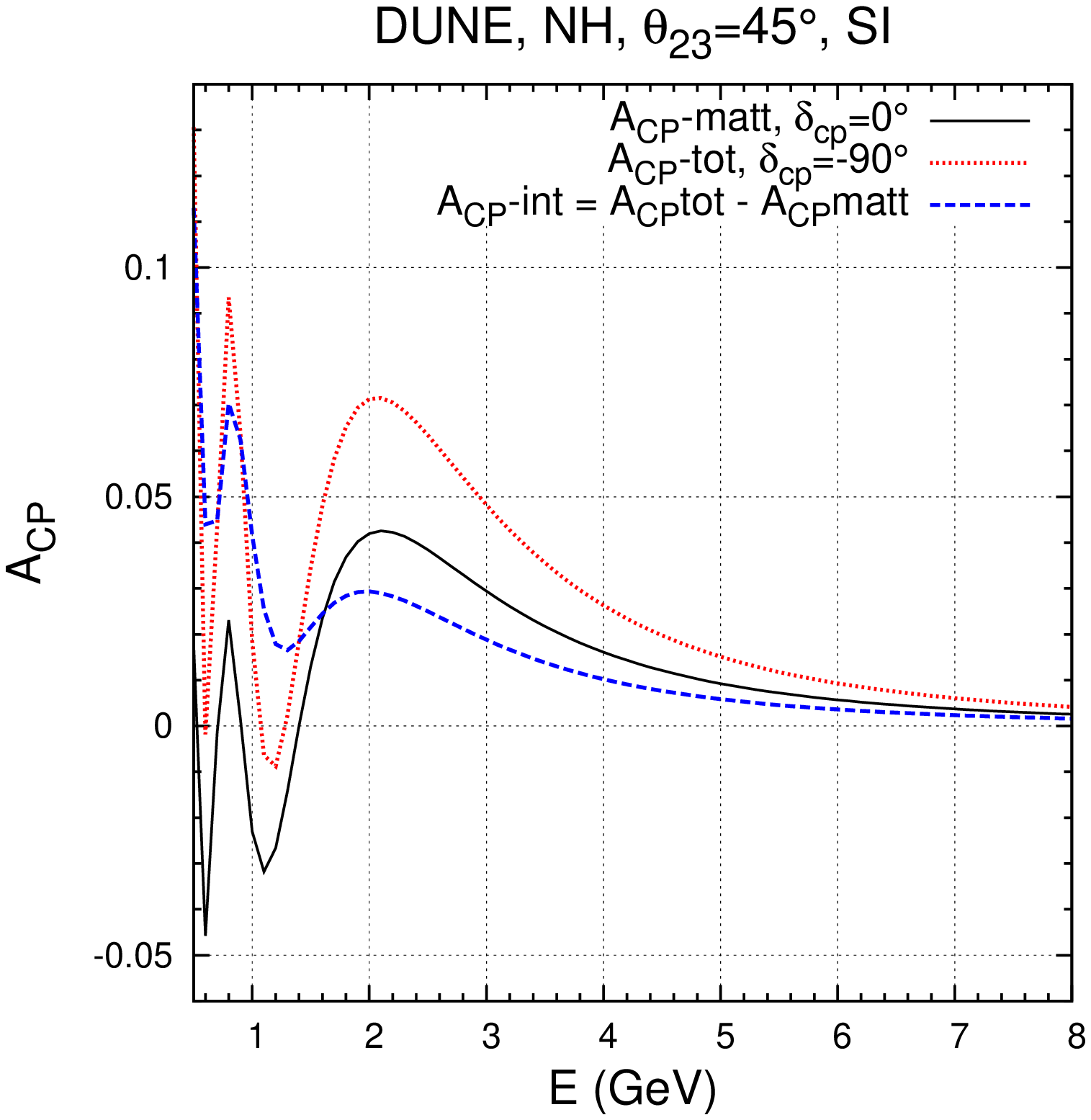} 
                \includegraphics[height=6cm,width=7cm]{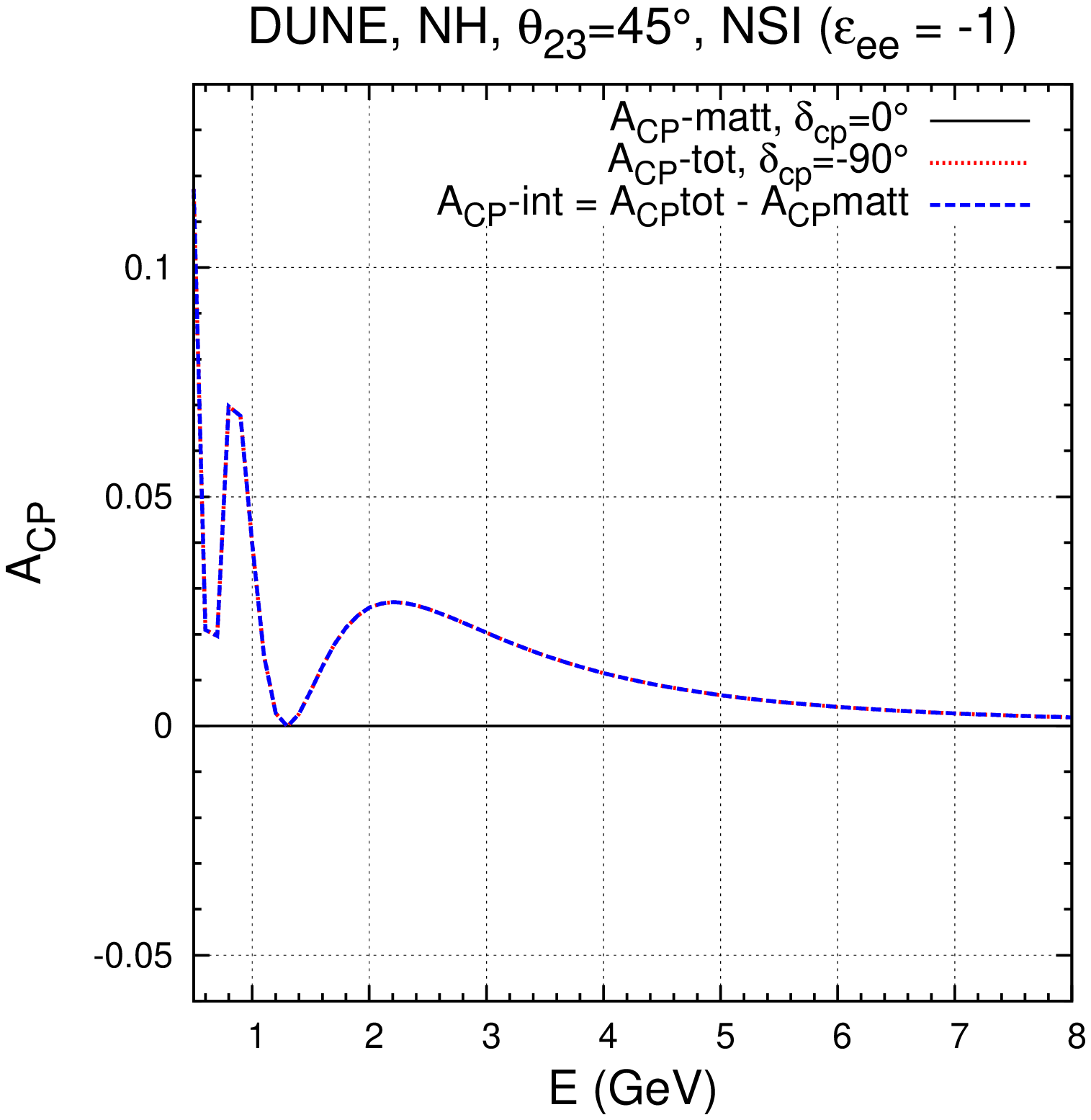}
        \end{tabular}
\vspace{-0.7cm}        
\caption{\footnotesize CP Asymmetry $A_{CP} = P_{\mu e} - \bar{P}_{\mu e}$ vs E(GeV) for DUNE considering standard paradigm in the left panel and NSI ($ \epsilon_{ee} = -1$) in the right panel.}
\label{fig:Asymm}
\end{figure}

Fig (\ref{fig:Asymm}) shows the CP asymmetry of DUNE defined as $A_{CP}= P_{\mu e} - \bar{P}_{\mu e}$ for NH and $\theta_{23}=45^\circ$. The experimentally measured asymmetry is $A_{CP}^{tot} = A_{CP}^{int} + A_{CP}^{matt}$. Here $A_{CP}^{int}= P_{\mu e}(\dcp) - \bar{P}_{\mu e}(\dcp)$ is the CP asymmetry quantifying the intrinsic Dirac CP phase $\dcp$ and $A_{CP}^{matt}= P_{\mu e}(\dcp^{\prime}) - \bar{P}_{\mu e}(\dcp^{\prime})$ quantifying fake CP phase $\dcp^{\prime}$ introduced by the asymmetry in the earth matter.  

In the left panel of fig (\ref{fig:Asymm}) the red (dotted) curve shows $A_{CP}^{tot}$ of DUNE when there is maximal 
CP violation with $\dcp = -90^\circ$. The black (solid) curve represents $A_{CP}^{matt}(\dcp^{\prime})$ which can be obtained in theory by taking $\dcp=0$. Now, the intrinsic CP phase can be quantified by plotting $A_{CP}^{int} = A_{CP}^{tot} - A_{CP}^{matt}$ as shown by the blue (dashed) curve. Clearly the intrinsic CP phase is different from the experimentally measured phase that is quantified by $A_{CP}^{tot}$.

The plot in the right panel of fig (\ref{fig:Asymm}) shows all the three asymmetries defined earlier
for  $\epsilon_{ee} = -1$.
Since $\epsilon_{ee} = -1$ nullifies the matter effect one can see from the black (solid) curve that the asymmetry induced by the earth matter $A_{CP}^{matt} = 0$. Thus, $A_{CP}^{tot} = A_{CP}^{int}$ which can be seen from the blue (dashed) and red (dotted) curves. That is in this special case, interestingly, the fake asymmetry introduced by the earth matter is nullified and the $A_{CP}^{tot}$ measured by the experiment is the same as $A_{CP}^{int}$. In this context it is 
worthwhile to study how the CPV sensitivity of DUNE is affected in this special case.


\section{Results and discussions}\label{sec:CPV sens}

In our numerical simulations we have used the General Long baseline Experiment Simulator (GLoBES) \cite{globes1} along with some additional packages \cite{globes-nsi1}. 
The experimental details are taken from \cite{Nath:2015kjg} except here, we have considered the detector volume to be 40 kt. The total $\chi^2$ in our analysis is defined by 
\be
\chi^2_{tot} = \chi^2_{stat} + \chi^2_{syst} + \chi^2_{prior}
\ee
where $\chi^2_{stat} = 2 \sum_i \lbrace N_i^{{\rm test}}-N_i^{{\rm true}}+N_i^{{\rm true}} ln \frac{N_i^{{\rm true}}}{N_i^{{\rm test}}} \rbrace$, $N_i^{{\rm true}}$
being the number of true events and $(N)_i^{{\rm test}}$ corresponding to the number of test events. The effect of systematics is included through the method of pulls. We have added a prior on $\sin^{2}2 \theta_{13}$ with an error of $\sigma(\sin^{2}2 \theta_{13}) = 0.005$. 

The true values that we have considered are, $\sin^{2} \theta_{12} = 0.297$, $\sin^{2}2 \theta_{13} = 0.085$, 
$\sin^{2} \theta_{23} = 0.5 $ (unless mentioned otherwise), $\dcp= - 90^{\circ}$, $ \Delta m^{2}_{21} = 7.37 \times 10^{-5} eV^{2}$ and $ \Delta m^{2}_{31} = 2.50 \times 10^{-3} eV^{2}$. These are consistent with the global analysis of neutrino oscillation data \cite{Capozzi:2016rtj,Esteban:2016qun,deSalas:2017kay}. 
We neglect the production and detection NSIs as these are bounded by an order of magnitude stronger than the matter NSIs \cite{Biggio:2009nt}.

\subsection{CPV sensitivity : Effect of non-zero NSI}
\begin{figure}[h!]
\begin{center}
 \begin{tabular}{lr}
\includegraphics[height=7cm,width=7cm]{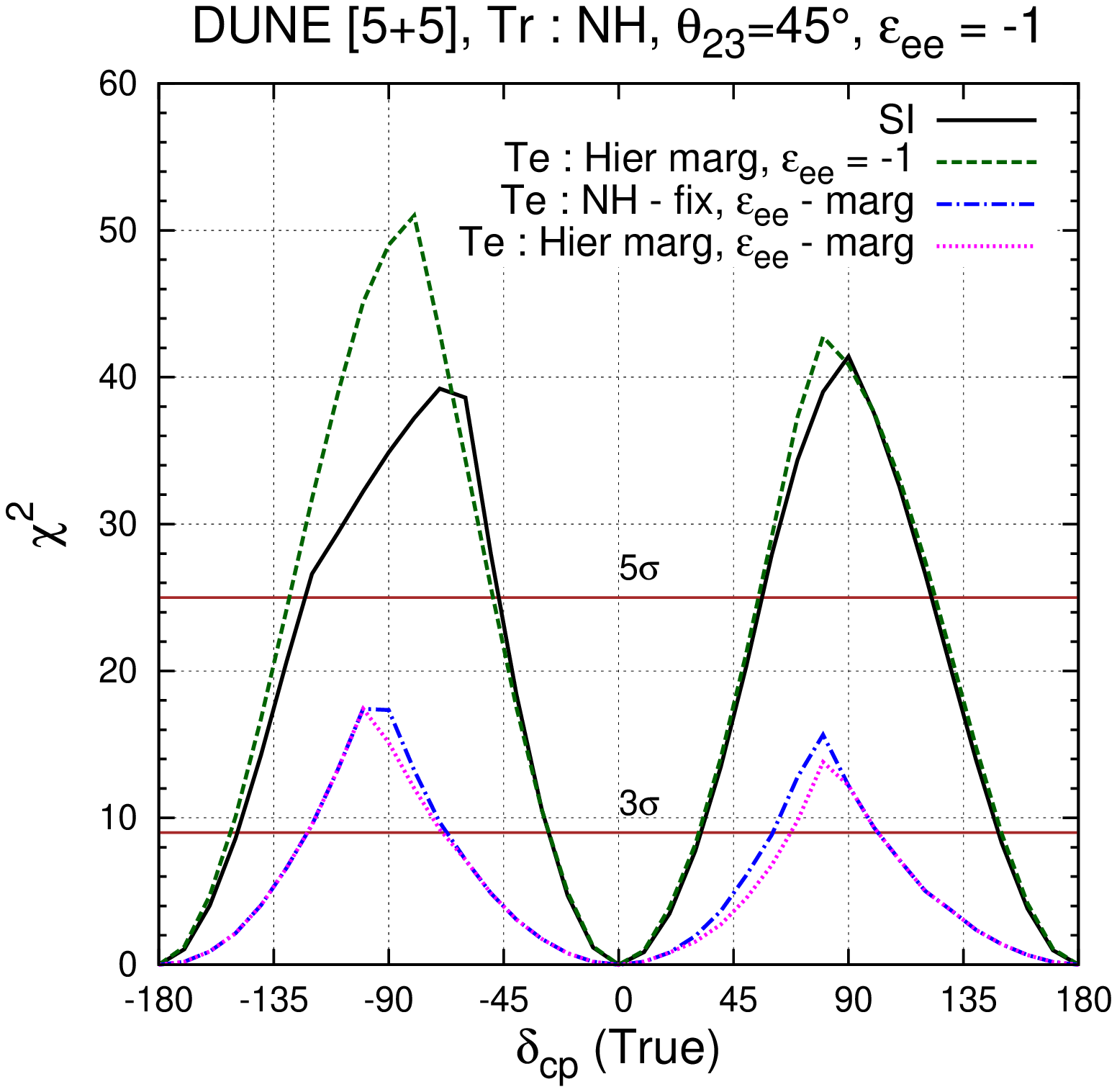}
\includegraphics[height=7cm,width=7cm]{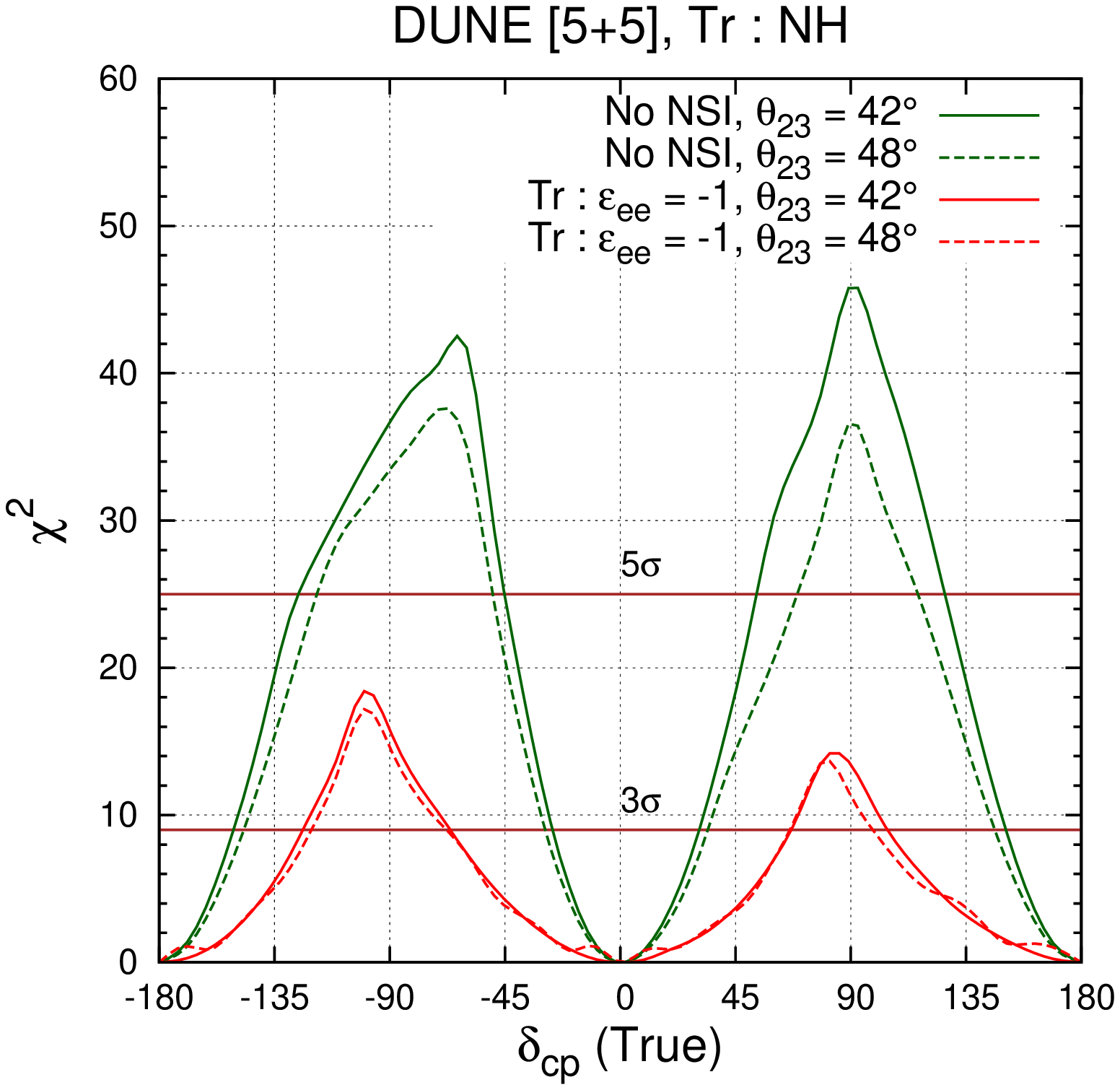}
 \end{tabular}
 \includegraphics[height=7cm,width=7cm]{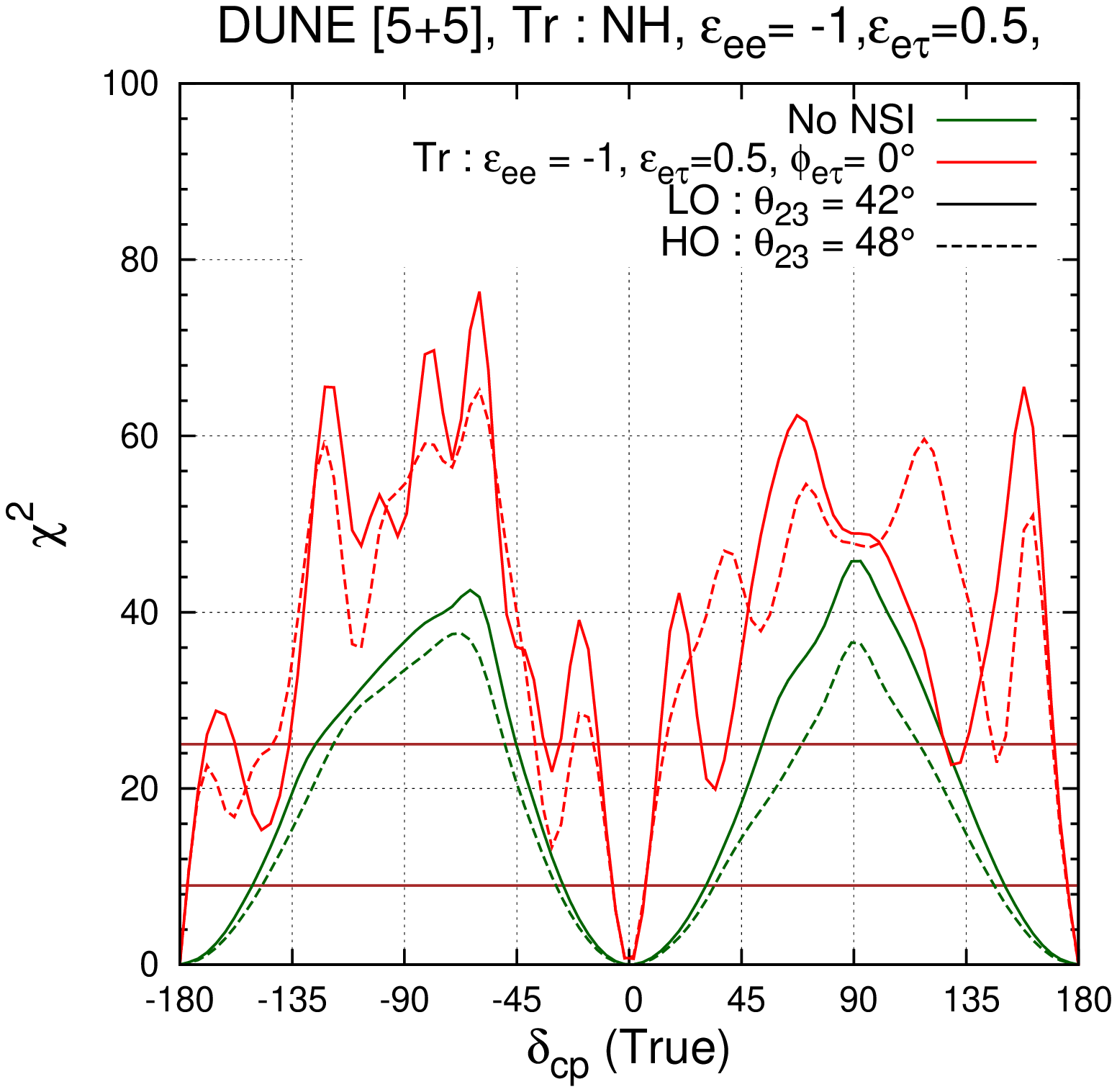}
 \end{center}
\caption{\footnotesize Upper Panel : CPV sensitivity of DUNE with $\epsilon_{ee} = -1$. Lower Panel : CPV sensitivity of DUNE with $\epsilon_{ee} = -1$, $\epsilon_{e \tau} = 0.5$, $\phi_{e \tau} = 0^\circ$ (see text for details).}
\label{fig:dune_cp_sens2}
\end{figure}

Fig.~(\ref{fig:dune_cp_sens2}) shows the significance with which CP violation, i.e. $\delta_{CP} \neq 0, \pm 180^\circ$ can be determined for different true values of $\delta_{CP}$. The CPV sensitivity is quantified by the test statistics $\chi^2$ which is defined as :
\be
\chi^2_{CPV} = min\left\lbrace \chi^2_{CP}(\delta_{CP}^{test}=0), \chi^2_{CP}(\delta_{CP}^{test}= \pi)\right\rbrace 
\ee
where $ \chi^2_{CP}$ is calculated by varying true $\dcp$ in the range $[-180^\circ,180^\circ]$.

We consider the true hierarchy to be NH and marginalized over the test hierarchy (hierarchy marginalization is done in all the plots unless otherwise mentioned), octant of $\theta_{23}$ and $\theta_{13}$.
The top panel of fig. (\ref{fig:dune_cp_sens2}) shows the CPV discovery potential of DUNE running for $(5\nu + 5\bar{\nu})$ years assuming $\epsilon_{ee} = -1$ in the data. In the test we have marginalized over $\epsilon_{ee} \in [-4,4]$ in addition
to the other parameters stated above. 
Here, the horizontal solid brown lines represent $3 \sigma$ and $5 \sigma$ C.L. as labelled in the figure. 
In the left panel of the first row, the black (solid) curve corresponds to the standard case i.e. assuming no NSI in both data and theory while for other curves we assume that the data has non-zero NSI with $\epsilon_{ee} = -1$.
 The green (dashed) curve corresponds to the case where $\epsilon_{ee} = -1$ is kept fixed in both true and test event spectra while the hierarchy is marginalized.  We see that this gives a higher sensitivity than the 
standard case around $\dcp \sim -90^\circ$. For other values of 
$\dcp$ the CP discovery potential is not affected much. 
Thus, the intrinsic degeneracy by itself does not impair the 
CP discovery capability of DUNE. 
However, when one marginalizes over $\epsilon_{ee}$ in it's  
model-independent range 
of $[-4,4]$, the sensitivity is compromised 
for $\dcp$ both in the upper ($0<\dcp<180^\circ$) and the 
lower ($-180^\circ < \dcp < 0$)  half-planes which can be seen from the 
blue (dot-dashed) and magenta (dotted) curves. 
This is due to the continuous degeneracy of the form
($\epsilon \rightarrow \epsilon^\prime$, $\dcp \rightarrow \dcp^\prime$). 
Note that the blue (dot-dashed) curve is for fixed hierarchy 
indicating that the continuous degeneracy can reduce the CPV discovery 
power of DUNE even for the same hierarchy. One can understand this from fig. (\ref{fig:WH-WCP}) by drawing a horizontal line to intersect $\epsilon_{ee}=-1$ (blue dot-dashed) curve and $\epsilon_{ee}=-2$ (brown dashed) curve at $\pm 180^\circ$ which corresponds to ($\epsilon \rightarrow \epsilon^\prime$, $\dcp \rightarrow \dcp^\prime$) degeneracy. 

In the right plot of the upper panel of fig. (\ref{fig:dune_cp_sens2}), the green solid and dashed lines represent the case where both data and theory are consistent with the standard paradigm while true $\theta_{23} = 42^\circ (LO)$ and $\theta_{23} = 48^\circ (HO)$ respectively. While the red solid (dashed) line shows CP violation sensitivity of DUNE assuming true non-zero diagonal NSI $\epsilon_{ee} = -1$ and $\theta_{23} = 42^\circ (LO)$ ($\theta_{23} = 48^\circ (HO)$). From the red curves it can be seen that the CPV sensitivity is compromised when compared to the standard case for all values of $\dcp$ both in the lower and upper half-planes. The sensitivity for $\theta_{23} = 42^\circ (LO)$ is only slightly greater than that of $\theta_{23} = 48^\circ (HO)$. Thus, one can conclude that there is very low correlation between $\epsilon_{ee}$ and the octant of $\theta_{23}$.

The lower panel of Fig. (\ref{fig:dune_cp_sens2}) shows CPV sensitivity of DUNE running for $(5\nu + 5\bar{\nu})$ years assuming a non-zero diagonal and off-diagonal NSI in the data. In the previous work \cite{Deepthi:2016erc} the authors have shown that in the presence of non-zero off-diagonal NSI parameter ($\epsilon_{e \tau}$) the intrinsic hierarchy degeneracy that occurs at $\epsilon_{ee} = -1$ and $\dcp = -90^\circ$ gets transported to a different value of $\epsilon_{ee}$. Thus, to check the impact of this shift in the intrinsic hierarchy degeneracy on the CPV discovery potential, we have considered the same representative values of $\epsilon_{ee} = -1$, $\epsilon_{e \tau} = 0.5$, $\phi_{e \tau} = 0^\circ$ chosen in \cite{Deepthi:2016erc} as our true values while we marginalize over these parameters in the test plane.
Here, the horizontal solid brown lines represent $3\sigma$ and $5\sigma$ C.L. as labelled in the figure. 
The green solid and dashed lines represent the case where both data and theory are consistent with the standard paradigm while true $\theta_{23} = 42^\circ (LO)$ and $\theta_{23} = 48^\circ (HO)$ respectively. While the red solid (dashed) line shows CP violation sensitivity of DUNE while assuming true non-zero NSI with $\theta_{23} = 42^\circ (LO)$ ($\theta_{23} = 48^\circ (HO)$). The wiggly pattern of the red lines indicate the presence of various degeneracies discussed in the Fig. (\ref{fig:WH-WCP}). One can see that for these values the CPV sensitivity of DUNE gets enhanced as compared to the standard sensitivity even for true $\phi_{e \tau}= 0^\circ$. Since the variation in the probability 
with $\dcp$ is higher in the presence of $\epsilon_{e \tau}$ (can be seen from fig. (\ref{fig:WH-WCP})), the 
CP discovery potential got improved.

\begin{figure}[h!]
\begin{center}
 \begin{tabular}{lr}
\includegraphics[height=7cm,width=7cm]{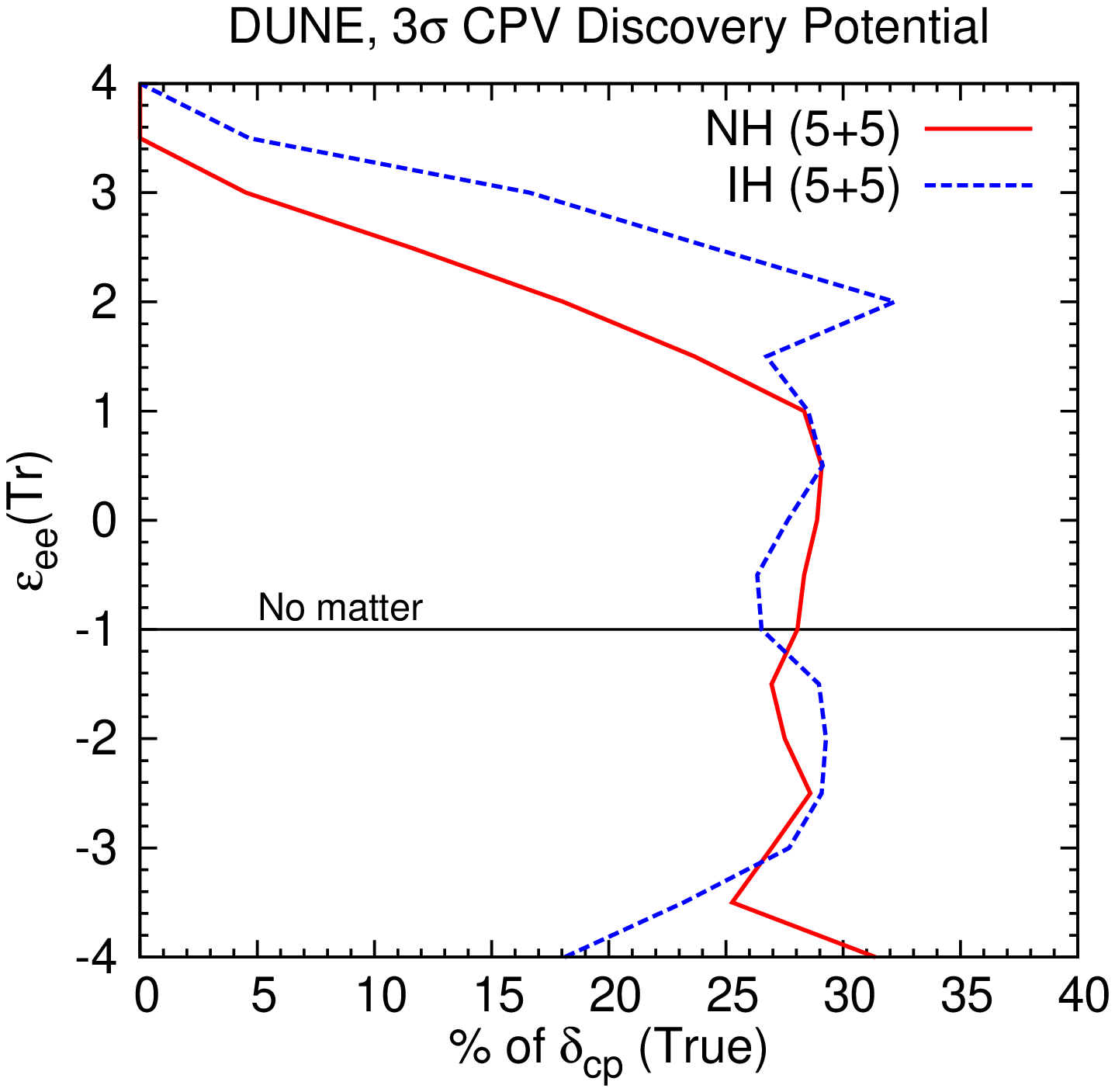} 
\includegraphics[height=7cm,width=7cm]{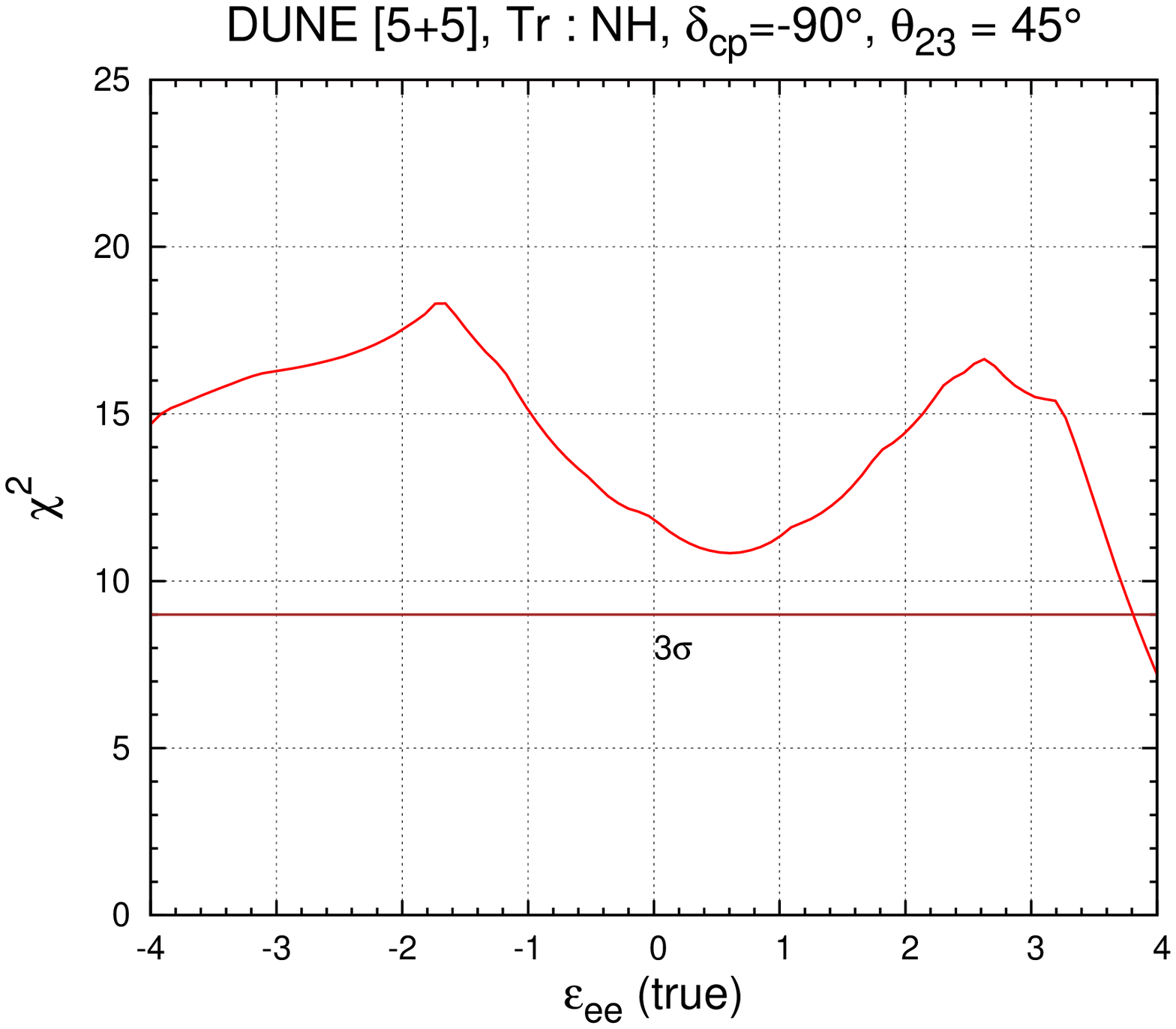}
\end{tabular}
 \end{center}
\caption{\footnotesize Left panel : Fraction of $\dcp$ for which CP violation can be discovered with $3\sigma$ significance at DUNE in the presence of non-zero $\epsilon_{ee}$. Right panel : The CP sensitivity as a function of true $\epsilon_{ee}$ at true $\dcp = -90^\circ$.}
\label{fig:chisq_epsee}
\end{figure}

In the above plots we have considered true $\epsilon_{ee} = -1$. To show how the CPV sensitivity depends on 
$\epsilon_{ee}$ we obtain the fraction of $\delta_{CP}$ values for which $3\sigma$ CPV discovery potential 
is possible by varying true $\epsilon_{ee} \in [-4,4]$.
CP fraction is one factor that can be used to quantify the effects of CP violation of a particular experiment.
This fraction determines the values of $\delta_{CP}$(true) for which the CP violation sensitivity can be 
obtained above a particular significance level. In left panel of fig. (\ref{fig:chisq_epsee}), we obtained CP fraction
of $\delta_{CP}$ values for which significance is above $3\sigma$ for different values of $\epsilon_{ee}$
in order to comprehend its complete effect.
From the red (solid) and blue (dashed) curves in the left panel of the fig. (\ref{fig:chisq_epsee}) it can be seen that CP violation discovery with $3\sigma$ significance can be obtained for $25\%$ of the $\dcp$ values when $\epsilon_{ee}$ varies from $\in [1,-3.8]$ for NH and $\in [2,-3.2]$ for IH. 

 As the global oscillation data hints $\dcp = -90^\circ$ we plot $\chi^2$ vs $\epsilon_{ee}$ in the right panel of fig. (\ref{fig:chisq_epsee}) for true $\delta_{CP} = -90^\circ$, $\theta_{23}=45^\circ$ and assuming NH. It can be seen from the fig that the CPV discovery potential remains above $3\sigma$ for almost the full range of $\epsilon_{ee}$. For $\epsilon_{ee} = 0$ the $\chi^2$ is $11.84$ when we use NSI to fit the data. 
However, if we compare with the standard case presented in figure 3 the $\chi^2$ is $34.87$. 
Thus the $\chi^2$ undergoes a large reduction in presence of NSI. 
And for no other values of $\epsilon_{ee}$ the $\chi^2$ goes as high as $34.87$
. Thus the 
effect of $\epsilon_{ee}$ is to reduce the CPV discovery potential of DUNE as compared to the standard case. This was also seen earlier in 
fig. \ref{fig:dune_cp_sens2} for $\epsilon_{ee} = -1$. 

\section{CP precision of DUNE}\label{sec:cp_precision}
Assuming the presence of diagonal NSI in nature, we check to what extent DUNE's CP measurement will get affected. To address this question we assume non-zero diagonal NSI $\epsilon_{ee} = -1$ along with a maximal CP violation $\delta_{CP} = -90^\circ$ in true scenario and in the fit we vary $\delta_{CP}$ in its full range $[-180^\circ, 180^\circ]$. 
We also show the results of CP precision of DUNE obtained by considering only standard neutrino framework both in test and true, as a benchmark for comparison. Since, the correlation between $\epsilon_{ee},\epsilon_{e \tau}$ and the octant of $\theta_{23}$ is very less henceforth we only plot the sensitivities corresponding to LO ($\theta_{23} = 42^\circ$).

\begin{figure}[h!]
 \begin{tabular}{lr}
\hspace{-0.5cm}
 \includegraphics[height=7cm,width=7cm]{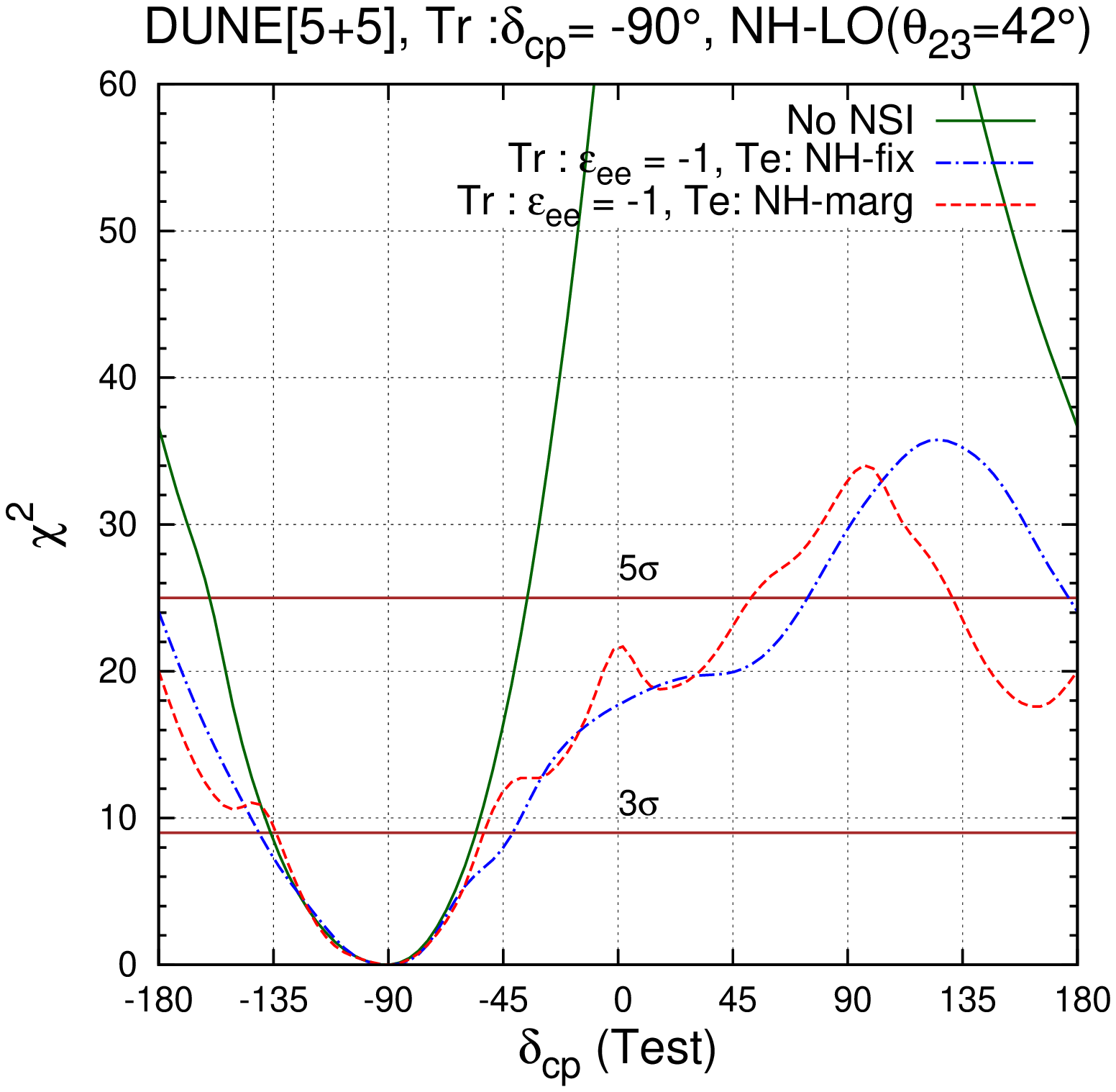}
 \includegraphics[height=7cm,width=7cm]{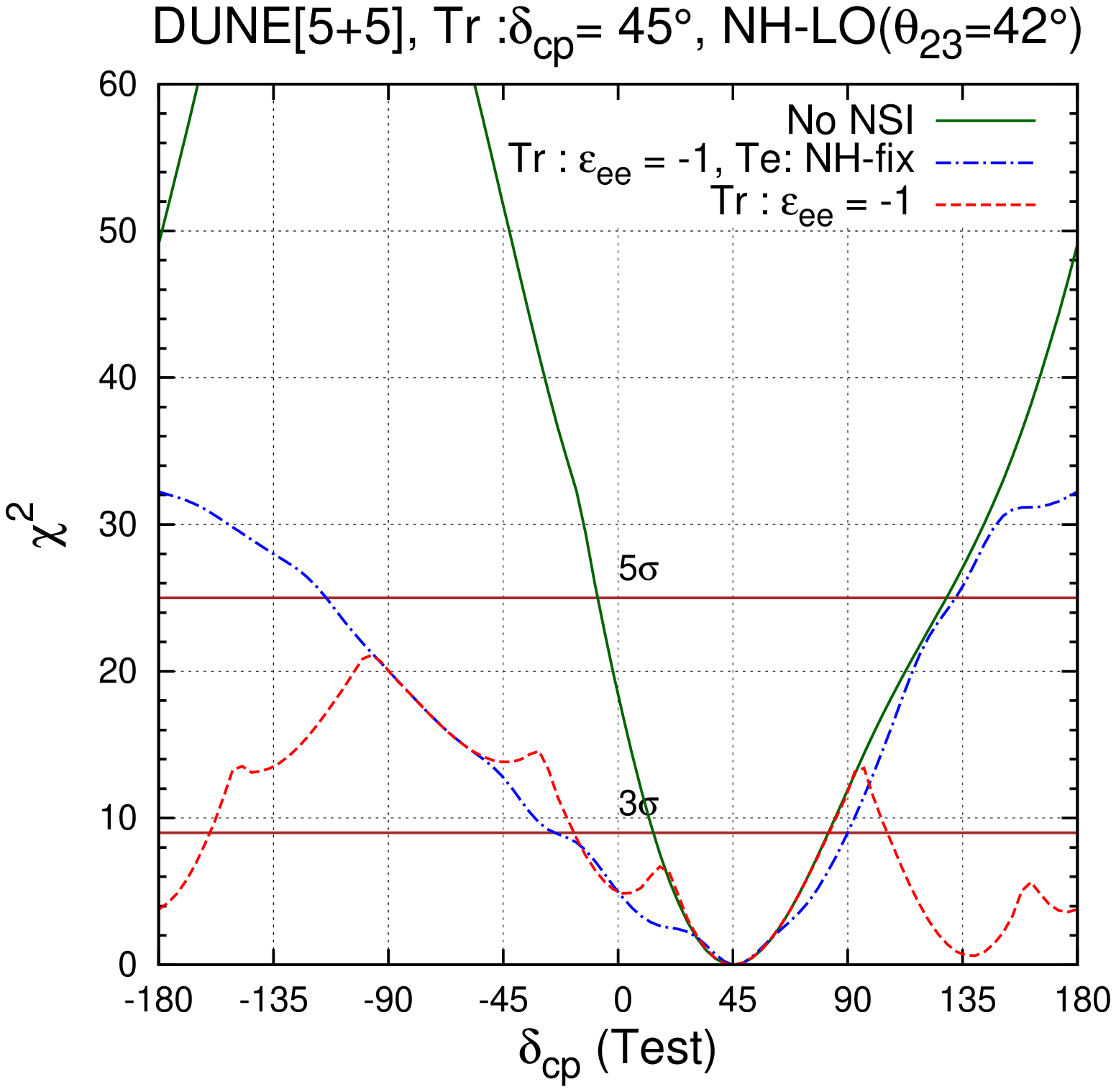}
 \end{tabular}
  \includegraphics[height=7cm,width=7cm]{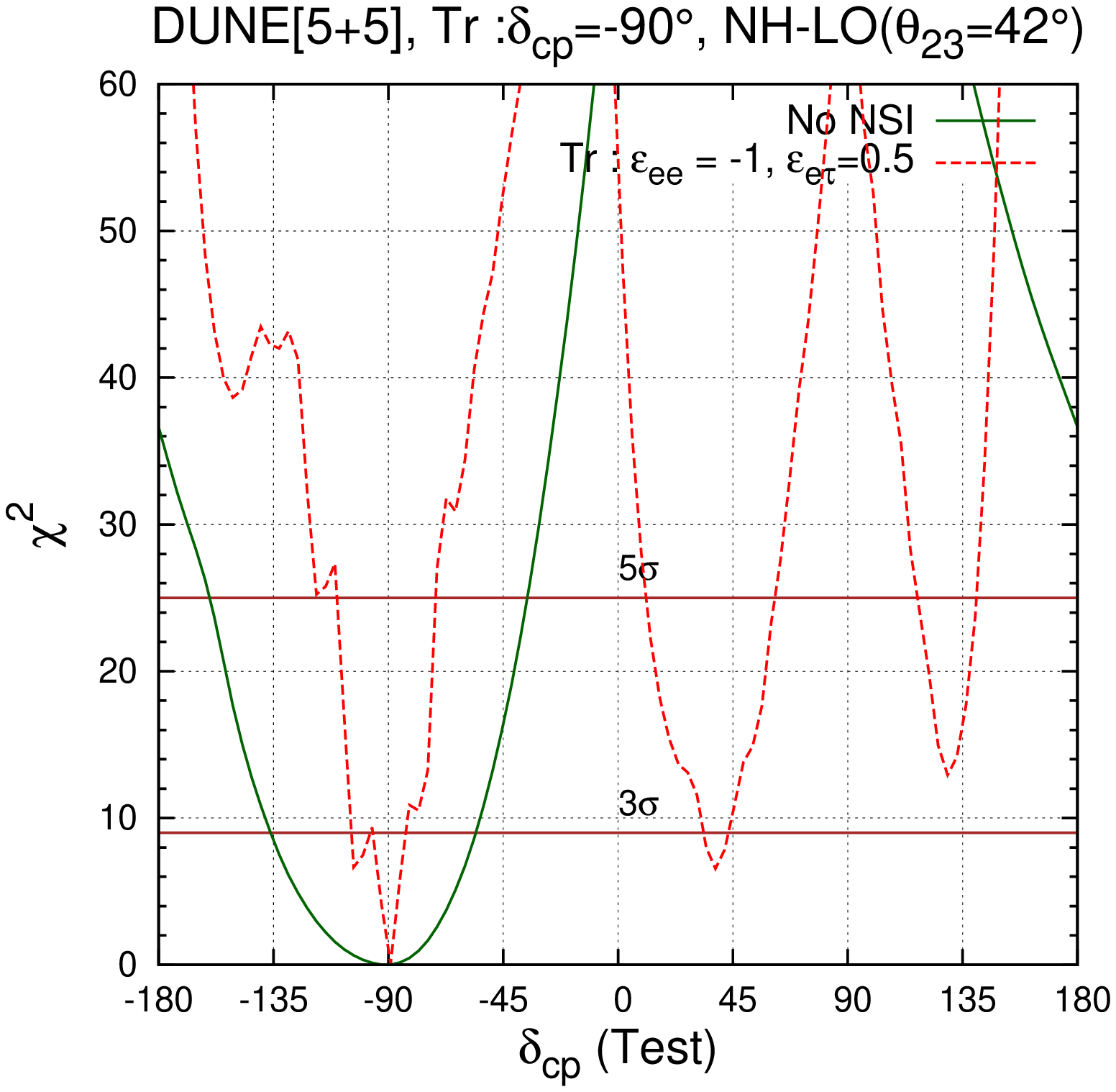}
\vspace{-4ex}        
\caption{\footnotesize  Upper panel : $ \chi^2  $  vs $ \delta_{CP} $ (test) assuming true hierarchy to be NH and $ \delta_{CP}$(true)$= -90^\circ$ ($45^\circ$) in the left(right) panel with true $\epsilon_{ee} = -1$.
Lower panel : $ \chi^2  $  vs $ \delta_{CP} $ (test) assuming true NH and $ \delta_{CP}$(true)$= -90^\circ$ with true $\epsilon_{ee} = -1$, $\epsilon_{e \tau} = 0.5$, $\phi_{e \tau} = 0^\circ$.}
\label{fig:dune_cp_prcsn_LOHO}
\end{figure}

Fig. (\ref{fig:dune_cp_prcsn_LOHO}) shows CP precision of DUNE in terms of $\chi^2$ vs test-$\delta_{CP}$ for true $\theta_{23} = 42^\circ (LO)$. While considering the true hierarchy to be normal, we have marginalized over $|\Delta m_{31}^2|$, $\theta_{13}$, $\theta_{23}$ and $\epsilon_{ee} \in [-4,4]$. We have also added gaussian prior to $\theta_{13}$.


The left plot in the upper panel of fig. (\ref{fig:dune_cp_prcsn_LOHO}) shows the CP precision of DUNE when true $\delta_{CP}=-90^\circ$. The green (solid) line corresponds to the standard case and the blue (dot-dashed) line corresponds to the case assuming $\epsilon_{ee} = -1$ in true and marginalizing over its full model independent range in the test while keeping hierarchy fixed to NH both in true and test planes respectively. The red (dashed) line corresponds to the same case as the blue (dot-dashed) curve except here the hierarchy is marginalized in the test.
For the SI case we see from the green curve that DUNE should be able to measure $\delta_{CP}$ with a precision of $\sim -90_{-45^\circ}^{\circ+36^\circ}$ at $3\sigma$ C.L. However when NSI exists, the blue (dot-dashed) and the red (dashed) curves reveal that the precision remain nearly the same at $3\sigma$ C.L. but become worse at $5\sigma$. This is because of the additional hierarchy degeneracies of the form $(\epsilon,\dcp) \rightarrow (\epsilon^\prime, \dcp^\prime)$ occurring between same as well as opposite hierarchy.

The right plot of fig. (\ref{fig:dune_cp_prcsn_LOHO}) corresponds to the CP precision of DUNE for true $\delta_{CP}=45^\circ$. For the standard case it can be seen from the green curve that DUNE has a 
CP precision of $\sim 45_{-27^\circ}^{\circ+45^\circ}$ at $3\sigma$ C.L. 
The blue (dot-dashed) and the red (dashed) curves are plotted with the same assumptions as in the left plot. It can be seen that the precision at $3\sigma$ and $5\sigma$ got worse because of the degeneracies of the form $(\epsilon,\dcp) \rightarrow (\epsilon^\prime, \dcp^\prime)$ occurring for the same hierarchy.
However when the hierarchy is marginalized, there appears another minima at $\delta_{CP} = 135^\circ$ corresponding to ($\dcp \rightarrow \pi - \dcp$ -- hierarchy) degeneracy which can be seen from the red (dashed) curve.
This shows that the CP precision of DUNE gets more compromised for any value of $\dcp$ other than $-90^\circ$ (here we show for $\dcp=45^\circ$) due to generalized hierarchy degeneracy which occurs for $\dcp \rightarrow \pi - \dcp$. 

The bottom panel of fig. (\ref{fig:dune_cp_prcsn_LOHO}) shows the CP precision of DUNE for true $\delta_{CP}=-90^\circ$ and in the presence of non-zero off-diagonal NSI $\epsilon_{e \tau} = 0.5$. It can be clearly seen that though the CP precision around the true value of $\dcp$ gets better, there are other local minima occurring because of the additional degeneracies introduced by the presence of non-zero $\epsilon_{e \tau}$. This can give multiple solutions in test $\epsilon_{e\tau} $ - test $\dcp$ plane.
\section{Conclusion:} \label{sec:conclusions}
In this paper we have studied the impact of the non-standard interactions 
on the CP sensitivity of DUNE. In particular, we have considered 
the impact of the diagonal NSI parameter $\epsilon_{ee}$. 
This is real and does not contain any extra complex phase. 
Nonetheless it can give rise to degeneracies of the form 
$(\epsilon,\dcp) \rightarrow (\epsilon^\prime, \dcp^\prime)$ between 
same as well as  opposite hierarchies.  
In particular, this parameter is responsible for 
the generalized hierarchy degeneracy 
$\epsilon_{ee} \rightarrow -\epsilon_{ee} - 2, \dcp \rightarrow \pi-\dcp$ 
which is known to adversely affect the hierarchy sensitvity of DUNE.  
This degeneracy can be resolved if the CP phase $\dcp$  or the 
diagonal NSI  parameter could be 
measured accurately. 
However  
for the special case of this with $\epsilon_{ee} = -1$ and $\dcp = \pm 90^{\circ}$ 
the  hierarchy would remain undetermined for all baselines and energies. 
This is known as the intrinsic hierarchy degeneracy. 
We have investigated how all these degeneracies can influence the 
CP discovery potential and precision measurement of $\dcp$ at DUNE. 
Taking the true value of $\epsilon_{ee} = -1$  
we obtain the CP discovery $\chi^2$  assuming 
presence of NSI in both data and theory.  We find that the 
CP discovery potential  gets negatively affected due to this 
parameter even when hierarchy is known. 
The degeneracy responsible for this is the continuous 
$(\epsilon,\dcp) \rightarrow (\epsilon^\prime, \dcp^\prime)$ degeneracy. 
However, if we keep $\epsilon_{ee}$ to be fixed as -1 in both  data and 
theory i.e do not marginalize over this parameter then the CP discovery 
potential can be  better than the standard case around $\dcp \sim -90^\circ$. 
Thus we conclude that the intrinsic hierarchy  degeneracy does not impair the 
CP discovery potential of DUNE. We have also shown that there is 
minuscule interplay between $\epsilon_{ee}$ and the octant of $\theta_{23}$.
We have further studied the impact of introducing a  
non-zero off-diagonal parameter $\epsilon_{e \tau}$   
in the analysis. We 
find that for the representative 
value considered by us the CP discovery potential is enhanced even 
when the true value of the complex phase associated with this parameter 
is taken to be zero. 
Additionally, we study the fate of CP discovery potential of DUNE for 
other true values of $\epsilon_{ee}$ by presenting the fraction 
of $\dcp$ values for which $3\sigma$ CP sensitivity can be attained 
as a function of this parameter. We find that this fraction is 25\% for 
$\epsilon_{ee} $ approximately between 1 and -3. 
We also study the maximal CPV discovery reach as a function of $\epsilon_{ee}$ 
for $\dcp = -90^\circ$, which is the best-fit value from global analysis 
of current world neutrino data.  In this case we find that although the 
CPV discovery $\chi^2$ stays above $3\sigma$ over almost the full allowed 
range of $\epsilon_{ee}$ this is less than the standard case.  
We have also studied the CP precision at DUNE. 
We find that for $\epsilon_{ee} = -1, \dcp = -90^\circ$ as true value 
the $\dcp$ precision at $3\sigma$ is comparable to the case with no NSI. 
However at $5\sigma$ the precision is worse. 
We have shown that this can again be attributed 
to $(\epsilon,\dcp) \rightarrow (\epsilon^\prime, \dcp^\prime)$ degeneracy which occurs 
for same and opposite hierarchies.
However if we take $\dcp(true)$ as different from $\pm 90^\circ$ 
the precision is compromised due to $\dcp \rightarrow \pi-\dcp$ 
occurring in the generalized hierarchy degeneracy. For a representative non-zero $\epsilon_{e \tau} = 0.5$ 
the CP precision around the true value improves but there are 
other local minima which can give spurious solutions. Further 
detailed study may be required to understand whether these degeneracies 
can be resolved by combining various neutrino oscillation experiments.

\textbf{Acknowledgement:} The research work of NN  was supported in part by the National Natural Science
Foundation of China under grant No.11775231.

\bibliography{nsi_neutrino}

\end{document}